\documentstyle[aps]{revtex}
\begin{document}
\input epsf 
\draft
\title{Domain Size Effects in Barkhausen Noise}
\author{M. Bahiana, Belita Koiller} 
\address{Instituto de F\'\i sica, UFRJ\\
	Rio de Janeiro, RJ, Brazil}
\author{S.L. A. de Queiroz}
\address{Instituto de F\'\i sica, UFF\\
	Niter\'oi, RJ, Brazil}
\author{J. C. Denardin,  R. L. Sommer}
\address{Departamento de F\'\i sica, UFSM\\
	Santa Maria, RS, Brazil}
\date{\today}
\maketitle
\begin{abstract}
The possible existence of self-organized criticality in Barkhausen noise is
investigated theoretically through a single interface model, and
experimentally 
from measurements in amorphous magnetostrictive ribbon Metglas 2605TCA under
stress. Contrary to previous interpretations in the literature, 
both simulation and experiment indicate that the presence of a cutoff
in the avalanche size distribution may be attributed to finite size effects. 
\end{abstract}
The Barkhausen effect consists of magnetic noise caused by 
erratic jumps in the magnetization
of a ferromagnetic material, under an increasing applied magnetic 
field \cite{barkorig}.
A simple explanation for this effect is the combination of random pinning
of domain 
walls by defects and the driving external field, which is essentially the
same mechanism
present in stick-slip processes \cite{feder1}.
Recently the statistical behavior of Barkhausen noise has attracted much
interest,  due to the possibility of providing 
an experimental realization of self-organized critical (SOC)
behavior\cite{weiss94,urbach,sethna1}.
The subject is controversial, however.
We concentrate here on the results obtained by  
Urbach {\em et al}, (UMM) \cite{urbach} and Perkovi\'c {\em et al} (PDS)
\cite{sethna1}. 

UMM measured the avalanche size probability
distribution function in an Fe-Ni-Co alloy, and found power-law
decay over approximately two decades, followed by an exponential cutoff.
The same result was also observed in numerical simulations of the
interface motion.
The power-law behavior, obtained without any intentional fine-tuning of
parameters, suggests
that this system self-organizes into a critical state. On the other hand,  
PDS argued that such behavior can be
explained without resource to SOC concepts: 
in their view, the power-law decay
followed by a cutoff is evidence that the system is {\em near} but not
quite {\em at} a conventional critical point.
They performed simulations for the random-field Ising model (RFIM) under an
external field, taking the local (pinning) fields to be gaussian-disordered 
with standard deviation $R$. The avalanche-size distribution is also
characterized by a power law followed by a cutoff, and the power-law regime
increases over several decades as $R$ approaches a critical disorder
$R_C$.  

Although UMM and PDS approach the problem with apparently similar
models, their conclusions regarding  the critical nature of the Barkhausen
noise
are in contradiction. 
Here we show that in reality, the  ingredients
used in either model differ in crucial aspects  where the
onset of SOC is concerned,
so it is not surprising that they
end up with different findings.

We investigate this question by using the simple model proposed by UMM
\cite{urbach} for the motion of a single domain wall in the
Barkhausen noise regime.
We find that the existence of a cutoff 
in the UMM model can be traced back to finite-size effects; 
experimental results, also to be described, bear out the idea that 
the cutoff to be found there originates from corresponding aspects in real 
systems.

In UMM's model, the interface at time $t$ is described,
in space dimensionality $d$, by its height 
$h(\vec{\rho_i},t)$, where $\vec{\rho_i}$ is the position-vector of 
site $i$  in a $(d-1)$-dimensional lattice. 
At each $t$, the height function $h_i = h(\vec{\rho_i},t)$ is assumed to be 
single-valued, 
so there are no overhangs on the interface. Thus  the 
interface element corresponding to the $d$-dimensional position-vector 
$\vec{r_i}=(\vec{\rho_i},h_i)$ may be unambiguously labelled by $i$. 
Simulations are performed on a $L^{d-1} \times \infty$  geometry,
with the interface motion set along the infinite direction.
Therefore finite-size effects are controlled by the length parameter $L$. 
Each element $i$ of the interface experiences a force of the form:
\begin{equation}
f_i=u(\vec{r_i})+\frac{k}{z}\left[\sum_{j=1}^z h_{\ell_j(i)}-z
h_i\right]+H_e~,
\label{force}
\end{equation}
where 
\begin{equation}
H_e=H-\eta M~.
\label{He}
\end{equation}
The first term on the RHS of (\ref{force}) represents the pinning force,
$u$, and brings quenched disorder into the model by being chosen 
randomly, for each lattice site $\vec{r_i}$,  from a
Gaussian distribution of zero mean and standard deviation $R$. Large negative
values of $u$ lead to local elements where the interface will tend to be
pinned, as described in the simulation procedure below.
The second term corresponds to a cooperative interaction among interface 
elements,  assumed here to be of elastic (surface tension)
type. In this term, $\ell_j(i)$  is the position of the $j$-th nearest 
neighbor of site $i$ and $z$ is the coordination number of the  
$(d-1)$-dimensional lattice  over which the interface projects.
The tendency of this term is to minimize 
height differences among interface sites: higher (lower) interface elements 
experience a negative (positive) force from their neighboring elements. 
The force constant $k$ gives the intensity of the elastic coupling, 
and is taken here as the unit for $f$.
The last term is the effective driving force, resulting from the applied 
uniform external field $H$ and a demagnetizing field which is taken to be 
proportional to 
$M=(1/L^{d-1})\sum^{L^{d-1}}_{i=1} h_i$, 
the magnetization of the previously flipped spins for a lattice of  width $L$.

Other models discussed in the literature\cite{sethna1,ji,stanley}
have the same basic ingredients 
described in (\ref{force}), namely  local quenched disorder, a cooperative
term and a driving field. 
In the RFIM, for example, the cooperative term is not elastic, but is driven 
by nearest-neighbor exchange interactions \cite{sethna1}.
Accordingly, power-law distributions 
are usually obtained along several decades of avalanche sizes; however,
the question of whether an SOC-like mechanism is present is
an altogether different matter. What is essential for SOC is that
no fine-tuning of parameters be required to  keep the system at a 
critical state. Such a distinction is clearly illustrated in 
Ref.~\onlinecite{urbach}, where several variants of the present model
were introduced. It was established that even without including
any demagnetizing effect at all, a power-law distribution of avalanche
sizes would arise if the external field were kept close to its
(sharply-defined)
critical value for interface depinning. Including a demagnetizing field
proportional to the local magnetization did away with fine-tuning: rather,
the effective field $H - \eta M$ was seen to stay approximately constant as
$H$
and consequently $M$ increased. Even then, the negative auto-correlation
between avalanche sizes at short times (observed in experiments, and also
believed to be an essential feature in the theory of SOC) only arose
in simulations when the {\it global} magnetization term $H_e$ shown in
(\ref{He}) was considered. 
Given the established adequacy of the UMM model to describe SOC-like aspects 
of experiments, we concentrate our simulations in that same model.

We start the simulation with a flat wall and zero applied field. 
All spins above it are unflipped. The force $f_i$ is then
calculated for each site, and each spin at a site with
$f_i\geq 0$ flips, causing the interface to  move up one step. 
The magnetization is updated, and this process continues until
$f_i<0$ for all sites, when the interface comes to a halt. 
The external field is then increased by the minimum amount needed to bring
the most weakly pinned  element to motion. The  avalanche size corresponds
to the number of spins flipped between two interface stops. 
The field value can be used as a time scale (the only relevant one in the 
case). This mimics the experimental setup of a linearly increasing field
during the data acquisition interval (one might as well use the accumulated
number of interface stops as a surrogate time scale; we did both, with
the same qualitative results).

\begin{figure}
\setlength{\unitlength}{1mm}
\begin{picture}(90,90)(0,0)
\put(50,5){\epsfxsize=8cm\epsfbox{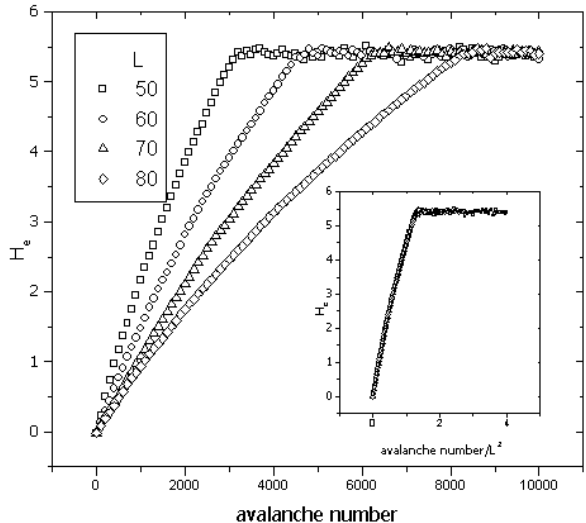}}
\end{picture}

\caption{Simulation on a $3-d$ lattice for systems with $R = 5.0$, $k = 1$,
$\eta = 0.05$ and  different widths. Starting the external field at $H=0$,
the effective field $H_e$ grows linearly and saturates at a critical value
after a transient. Here we see this behavior as a function of the avalanche
number and, in the inset,the same but with avalanche number scaled by
$L^2$.}
\label{fig-trans3d} 
\end{figure}

Our simulations have been conducted in $d=2$ and $d=3$, in square and 
cubic space lattices respectively. 
Here we concentrate on the 3-d results.
After a transient, the effective field always settles onto a critical value 
$H_c$ which depends on $R$, $k$ and $\eta$.
The fact that the system tunes itself to a
constant effective field is in itself an indication of SOC-like behavior. 
If the external field is started at a higher value, a large avalanche occurs
and brings the effective field back to the adequate value. On the other
hand, if the field
starts from zero there is a transient corresponding to a series of 
small avalanches. We find that the number of avalanches in this transient is
proportional to $L^{d-1}$. This means that an infinite system would need an
infinite number  of avalanches to reach criticality. As an illustration of this
behavior, we present in Figure
(\ref{fig-trans3d})
the effective field in a $3-d$ system for $R=5.0$, $k =1$
and $\eta = 0.05$ 
 for different simulation cell sizes $L$.
\newpage
\vspace*{.1cm}

\begin{figure}
\setlength{\unitlength}{1mm}
\begin{picture}(90,60)(0,0)
\put(50,5){\epsfxsize=8cm\epsfbox{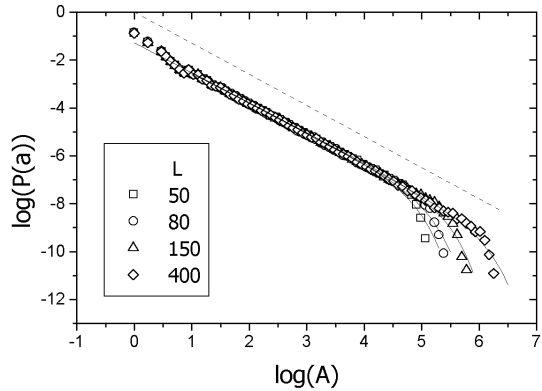}}
\end{picture}

\caption{Avalanche size distribution calculated from simulations 
for $R$=5.0, $k$= 1, $\eta = 0.05$ and the indicated values of lattice widths
$L$, in $3-d$. 
Thin solid lines are fits to the form $P(A)\propto A^{-\alpha}\exp
(-A/A_0)$
with  $\alpha$ between 1.23 and 1.35 and $A_0$, the cutoff size, 
increasing as a power of  $L$ (see text).
The dashed line has  slope $-1.3$, for comparison with UMM's experimental
data.
}
\label{fig-hist} 
\end{figure}

The calculated avalanche-size distribution 
corresponds to a power-law with a cutoff, as obtained by Urbach 
{\em et al} in
Barkhausen experiments and in simulations \cite{urbach}, and also by
Perkovi\'c {\em et al} \cite{sethna1} within their RFIM model. 
It was shown in the latter reference that the cutoff is intrinsic to
the RFIM, if the system is away from the critical point $R_C$.
In what follows we present evidence that the nature of the cutoff in the
Barkhausen effect, both in simulations of the Urbach model
 and in experiments, is a finite-size effect. 
The characterization of the cutoff
as a finite-size effect was recently suggested by Narayan\cite{narayan}
through the analysis
of a continuum model closely based on UMM's, though no attempt was
made there to quantify the relationship.
We have examined the dependence of the cutoff on the simulation cell size
$L$ by collecting  series of 100,000 avalanches for $L$ = 50,80,150,200 and
400.
Figure \ref{fig-hist} shows the avalanche size distribution for some values
of $L$ in $3-d$ lattices. 
The transient was eliminated by starting
the external field
near $H_c$. We can clearly see that the cutoff increases with $L$.
The histograms were fitted by the function $P(A)\propto
A^{-\alpha}\exp (-A/A_0)$ with  $\alpha$ in the range $(1.23\pm 0.02
-1.35\pm 0.02)$. The parameter $A_0$, which defines the cutoff size, is
strongly 
dependent on $L$: $A_0\propto L^{1.4\pm 0.1}$. 
We have also varied the disorder parameter $R$, in an attempt to find
an effect similar to that described by PDS in their RFIM model. However,
the final picture was qualitatively always the same as shown in 
Figure \ref{fig-hist}, with roughly the same power-law dependence of $A_0$ on
$L$.
In $d=2$ we used lattices of width $L =$ 100, 500, 1000, 2000, 3000 and
5000 from which
a value of $\alpha = 0.83\pm 0.08-1.03\pm 0.03$ and $A_0\propto L^{0.78\pm
0.06}$ were
obtained.
Thus, we conclude that the presence of a cutoff
in the avalanche size histogram is a finite-size effect.
Though it may seem puzzling that the finite {\it transverse} dimensions
can influence the characeristics of interface motion along the {\it unbound}
direction of growth, a qualitative picture of the corresponding mechanism
is as follows.
Each  time the $(d-1)-$ dimensional lattice is swept, the number
of distinct chances for the interface to move is of order $N \sim L^{d-1}$;
thus,
if the typical probability for a given interface element not to advance
(that is, to have $f_i <0$) is
$p$, the interface as a whole will come to a halt, marking the
end of an avalanche, only if $f_i <0$ for {\it all} elements. This happens
with probability $p^N$, hence for large $L$ larger avalanches become less
unlikely.
\vspace*{-.1cm}

The above result clearly demonstrates the different nature of the cutoff in
the UMM and PDS simulation models.  
Both models describe domain walls advancing in a disordered magnetic
medium: The key difference between them is the presence of a demagnetizing
field, proportional to the growing domain magnetization. This field is an
essential element to bring the simulations into an SOC regime.

Although variations in $L$ are easily accessible in simulations, this is 
not a trivial parameter to vary in experiments. For magnetic systems,
one would expect $L$ to be related to the typical magnetic domain size in
a sample. 
Magnetic materials have the interesting property of magnetostricition, which
is the change in internal domain configurations
as a  response to applied anisotropic stress.  Positive magnetostrictive
materials show an 
increase in internal domain wall lengths under applied stress
\cite{yamamoto}, and may
therefore
be employed in investigating experimentally $L$-dependent effects 
in  Barkhausen noise experiments. 
We performed measurements in a disorderd material with this property, 
namely amorphous magnetostrictive ribbon Metglas 2605TCA under stress
$\sigma$. 
Different stress values were applied in order to increase
domain wall length, as seen in ref. \cite{yamamoto}.  Stress is expected
also to change the domain wall thickness  $\delta=\sqrt{{\cal A}/K_\sigma}$, 
where
${\cal A}$ is the effective exchange constant, $K_\sigma=\frac{3}{2}
\lambda_S \cdot \sigma $ and $\lambda_S = 27 \cdot 10^{-6}$ is the saturation
magnetostriction constant for this material. In amorphous materials, 
fluctuations in the stress due to local composition fluctuations generate the
pinning sites for domain walls. Their  magnitudes are much larger than those
originated from the external forces. Thus, the effect of the applied
stresses will be to order the domain wall structure by changing the domain
size and length. The magnetoelastic anisotropy, on the other hand, tends to
stretch existing domain walls \cite{yamamoto}.

Metglas 2605TCA samples (80 mm $\times$ 1 mm $\times$ 30 $\mu$m) were
pre-annealed in an Ar flow at $300^o$C for 15 min. in order to decrease the
stress level associated to the fabrication process. The measurements were
performed in an open magnetic circuit. As a consequence, there is a global
effective field acting on the domain walls; also, the average magnetization
rate $\dot{M}(t)$ and differential susceptibility were kept constant. The
samples were cycled in their hysteresis loops, excited by a slowly varying
(triangular, 0.2 Hz) field. The Barkhausen signal was detected by a small
($\approx$ 5 mm) coil wound around the central part of the sample. The signal
was preamplified by a SR550 low pass amplifier and digitized by a TDA320
oscilloscope. The low pass amplifier was set to an upper frequency limit
equal to half the sampling frequency. The waveform generator, current
source and preamp were fed by batteries in order to increase the signal to
noise ratio. At each cycle and starting from a given value of a trigger
field, a time series was acquired and stored for further processing. For
each stress level, we identify an avalanche with a jump in the voltage
level $V$.
As in Ref.\cite{urbach}, a threshold voltage was established according to
experimental resolution. This threshold defines the low-end cutoff of the
avalanche distribution, and shall not concern us particularly here as it
is the high-end of the distribution that will be sensitive to finite-size 
effects.   
It is to be noted, however, that especially for low-stress  data, the
combination of experimental resolution available and the intrinsic
characteristics of Metglas resulted in a rather narrow power-law
region. This in turn yielded quite a large spread in the fitted values
of the effective power $\alpha$, as seen below.

\begin{figure}
\setlength{\unitlength}{1mm}
\begin{picture}(90,60)(0,0)
\put(50,5){\epsfxsize=8cm\epsfbox{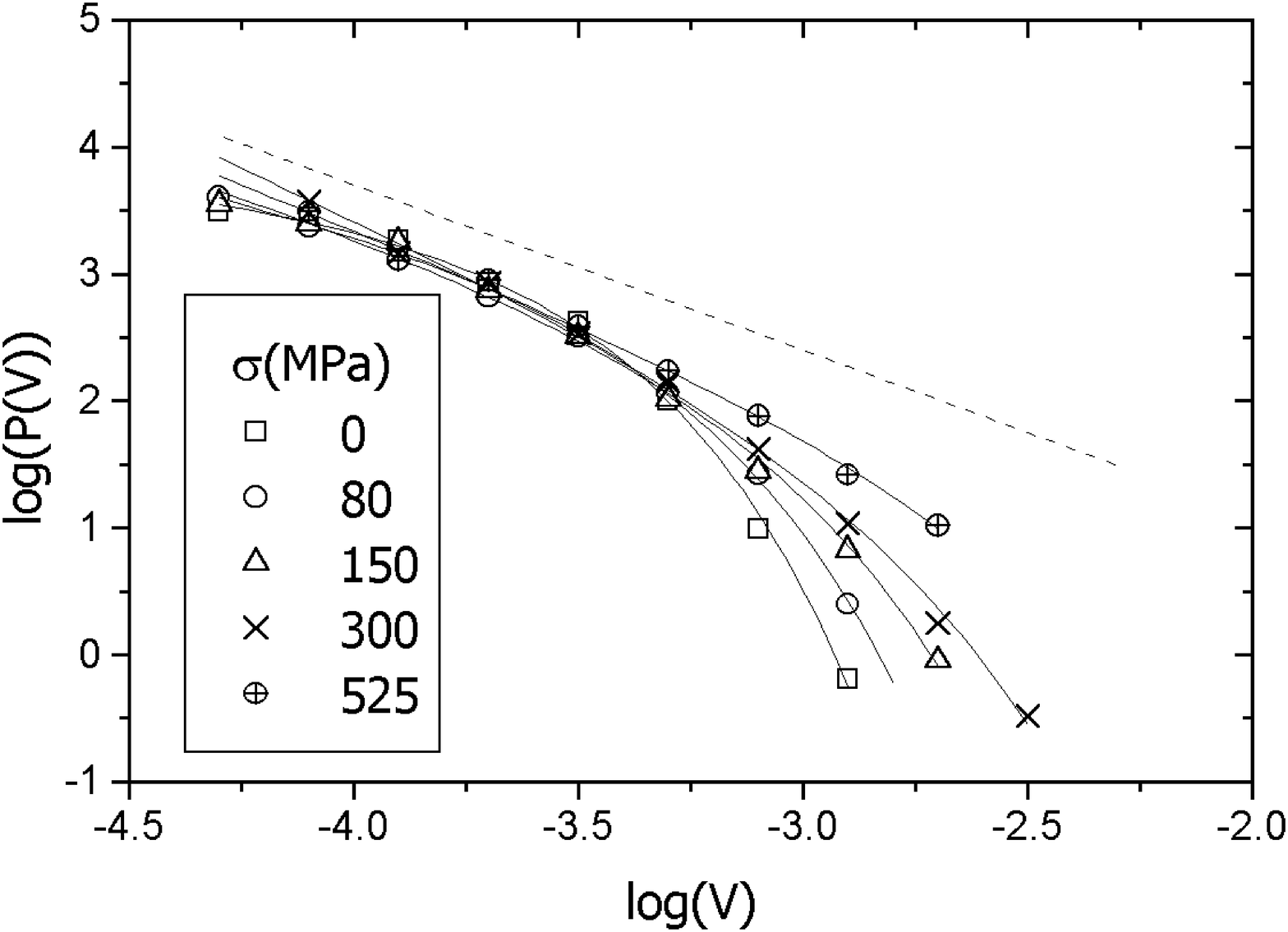}}
\end{picture}
\caption{Histogram of voltage jumps for different values of applied 
stress $\sigma$.
Thin solid lines are fits to the form $P(V)\propto V^{-\alpha}\exp
(-V/V_0)$ with 
$\alpha$ between 0.29 and 1.6 and $V_0$, the cut-off voltage, increasing 
as a power of  $\sigma$ (see text).
The dashed line, included for comparison with Fig.2 and with
UMM's experimental data, has the same slope as
the dashed line there, $-1.3$ .}
\label{histexp}
\end{figure}

The avalanche (voltage jump) size distributions were
obtained for $\sigma$ = 0, 17, 30, 80, 100,
150, 180, 230, 300, 400 and 525 MPa. Some of these results are shown in
Fig. \ref{histexp}.
\vspace*{.1cm}

It is clear that the
cutoff increases with the applied stress. Since stress increases the
magnetic domains, what we see here is again the dependence of the cutoff on
a typical domain size, in accordance with the simulations results. The
experimental histograms where also fitted with the function $P(V)\propto
V^{-\alpha}\exp (-V/V_0)$ giving $\alpha$ in the range $0.29\pm 0.1-
1.6\pm 0.1$ and $V_0\propto \sigma^{(0.65\pm 0.04)}$.
The fact that experimental values of $\alpha$  fall in a similar 
range to those from the $3-d$ simulations is both to be expected and in line
with the earlier results of UMM. On the other hand, 
the power that governs the dependence of $A_0$ on $\sigma$ need not coincide
with that which relates $A_0$ to $L$ in the numerical work.
In order to
predict a relationship between the two quantities, one would need 
to work out the connection of 
the physical mechanisms driving the interplay between finite
transverse dimensions and avalanche sizes, in the interface model, to 
the corresponding ones between applied stress and domain wall length in
actual samples. Thus far, we have not been able to do so.

From the above results we conclude that the cutoff in the avalanche size
histogram in Barkhausen systems is a finite size effect. This, together
with the presence of a self-tuning effective field and negative time
correlation for short times, is a strong indication that SOC is present.
Also, we find that any attempt to model Barkhausen noise must
necessarily include the demagnetizing field, for this is the key ingredient
for the above mentioned self-tuning.

\acknowledgements
This work was partially supported by CNPq, CAPES and FAPERGS(Brazil).
We thank M. Novak, J. Urbach and A. Hansen for interesting discussions.


%
%
\end{document}